\documentclass[journal]{IEEEtran}
\usepackage{amsfonts}
\usepackage{latexsym}
\usepackage{newlfont}
\usepackage{graphicx}
\usepackage{mathtools}
\usepackage{booktabs}
\usepackage{tikz}
\usetikzlibrary{snakes}
\usepackage[ruled,vlined]{algorithm2e}
\usepackage{algorithmic}
\usepackage{stfloats}
\usepackage{subfig}

\newcommand{\bn}{\mathbf{n}}
\newcommand{\bx}{\mathbf{x}}
\newcommand{\by}{\mathbf{y}}
\newcommand{\bw}{\mathbf{w}}

\newcommand{\bA}{\mathbf{A}}

\newcommand{\bd}{\mathbf{d}}
\newcommand{\be}{\mathbf{e}}

\newcommand{\bb}{\mathbf{b}}
\newcommand{\bD}{\mathbf{D}}

\newcommand{\bH}{\mathbf{H}}

\newcommand{\cR}{\mathcal{R}}
\newcommand{\br}{\mathbf{r}}

\newcommand{\bs}{\mathbf{s}}

\newcommand{\bI}{\mathbf{I}}

\newcommand{\bW}{\mathbf{W}}
\newcommand{\bv}{\mathbf{v}}

\newcommand{\bz}{\mathbf{z}}
\newcommand{\bZ}{\mathbf{Z}}

\newcommand{\bOmega}{\mathbf{\Omega}}
\newcommand{\bzero}{\mathbf{0}}
\newcommand{\diag}{\textrm{diag}}

\newcommand{\bdiag}{\textrm{block-Diag}}

\newcommand{\var}{{\text{Var}\,}}
\newcommand{\bdFT}{\check{\mathbf{d}}}
\newcommand{\brFT}{\check{\mathbf{r}}}
\newcommand{\bRFT}{\check{\mathbf{R}}}
\newcommand{\bwFT}{\check{\mathbf{w}}}

\newcommand{\argmin}{\operatornamewithlimits{argmin}}

\usepackage{ifpdf}

\usepackage{cite}
\usepackage{array}

\usepackage{mdwmath}
\usepackage{mdwtab}

\usepackage{eqparbox}
\usepackage{url}

\usepackage{color}

\begin{document}

%
\title{Sparse multichannel blind deconvolution of seismic data via spectral projected-gradient}
%
%
%

\author{Naveed~Iqbal, Entao~Liu, James~H.~McClellan,~\IEEEmembership{Life Fellow,~IEEE,}
        and~Abdullatif~A.~Al-Shuhail 
         \thanks{Naveed Iqbal and Abdullatif A. Al-Shuhail are with Center for energy and Geo processing (CeGP) at King Fahd University of Petroleum and Minerals. email: \{naveediqbal, ashuhail\}@kfupm.edu.sa.}
        \thanks{Entao Liu and James H. McClellan are with CeGP at Georgia Institute of Technology. email: \{entao.liu, jim.mcclellan\}@gatech.edu.}
}
\maketitle

\begin{abstract}
        In this work, an efficient numerical scheme is presented for seismic blind deconvolution in a multichannel scenario. The proposed method iterate with two steps: first, wavelet estimation across all channels and second, refinement of the reflectivity estimate simultaneously in all channels using sparse deconvolution. 
        The reflectivity update step is formulated as a basis pursuit denoising problem and a sparse solution is obtained with the spectral projected-gradient 
        algorithm\,--\,faithfulness to the recorded traces is constrained by the measured noise level.
        Wavelet re-estimation has a closed form solution when performed in the frequency domain by finding the minimum energy wavelet common to all channels.
        Nothing is assumed known about the wavelet apart from its time duration.
        In tests with both synthetic and real data, the method
        yields sparse reflectivity series and stable wavelet estimates results compared to existing methods with significantly less computational effort.
\end{abstract}


%
\IEEEpeerreviewmaketitle
\section{Introduction}
Seismic deconvolution is a standard procedure in seismic data processing in which the effects of a source wavelet are removed as much as possible \cite{Rie1997}, which also attenuates reverberations and short-period multiples. 
Deconvolution techniques are widely used in seismic exploration \cite{TrivdBAO2007,vdBPham2008} and seismology applications \cite{BosSac1997,Bos2004,BehShe2014}. 
Knowledge of the source wavelet is key, so a challenging problem in seismic deconvolution is \textit{blind deconvolution} where the blurring kernel, i.e., the seismic source wavelet, is unknown and must be estimated \cite{UlrVelAO1995}.
Recent work on multichannel semi-blind deconvolution (MSBD) \cite{MirCoh2017} addresses the situation where there is uncertainty in the assumed wavelet. Other semi-blind methods include the  $\phi_{HL}$ regularization based method \cite{Zhu2015}  and the Least trimmed squares (LTS) regularization based method \cite{Deng2014} that preserve the spectral details.

In a seismic survey, the convolution of the source wavelet with a subsurface reflectivity series is recorded as a seismic trace. 
In a multichannel scenario \cite{XuLiuAO1995,KaaTax1998,RamCohAO2010,MirCoh2017}, the seismic traces are typically modeled as convolutions of the same waveform with multiple reflectivity models. 
Early work on seismic blind deconvolution depended on two major assumptions: the impulse response of the earth is a white sequence and the source wavelet is minimum phase. 
In order to overcome these two limitations, homomorphic deconvolution \cite{OtiSmi1977,HervdB2012} and minimum entropy deconvolution (MED) \cite{Wig1978} were developed. 
In seismic applications, conventional multichannel methods cannot be applied directly. 
The major cause is the great similarity between neighboring reflectivity sequences, which makes the problem either numerically sensitive or, at worst, ill-posed and impossible to solve \cite{NosTak2016}. 

In recent attempts to tackle this issue, a sparsity promoting regularization approach has been proposed by \cite{GhoSac2012}. 
 Newer methods, called sparse multichannel blind deconvolution (SMBD) \cite{KazSac2014} and its variant modified SMBD \cite{KazGhoAO2016}, have been shown to perform well for both synthetic and real data sets. 
In \cite{Kazemi2016} authors used SMBD  to estimate source and receiver wavelets. However,  computational complexity of these  SMBD methods is proportional to the square of the number of traces which constrains them from being applied to large data sets directly. A common solution for this issue is to apply these algorithms on data patches. 
Another alternative is a series of variants of the deconvolution filtering design method, such as the widely adopted predictive deconvolution \cite{Yil2001}, and the fast algorithm for sparse multichannel blind deconvolution (F-SMBD) \cite{NosTak2016,Nose-Filho2018}. 
These approaches construct a deconvolution filter according to some criteria such as least-squares, or the smoothed $\ell_1$ norm of the deconvolved signal. 
Compared with the SMBD methods, their computation time is much lower, but they tend to produce a bandlimited deconvolution result that is not spiky enough.

In this work we propose an iterative blind deconvolution scheme with two phases: reflectivity series estimation based on straightforward basis pursuit denoising, and least-squares wavelet estimation that takes advantages of the common wavelet present in all traces across a multichannel seismic section.
The only assumption needed for the wavelet is that its time duration is limited, and known.
The wavelet estimation phase benefits from the bandlimited nature expected for a seismic source wavelet, but that is not an assumption.
In particular, we note that a very recent semi-blind method \cite{MirCoh2017}, which assumes the true wavelet belongs to a known dictionary of possible finite-length wavelets, can be solved by the proposed method without the need for a dictionary\,--\,known duration is sufficient.

The basis pursuit algorithm used here is the spectral projected-gradient (SPG) algorithm, which converges quickly, and suitable for large scale problems. 
More importantly, using SPG mitigates the tricky issue of choosing good regularization parameters, using instead a constraint based on noise power which can be measured.
In methods using regularization, the determination of a good $\lambda$ parameter is crucial to control the balance between sparsity of the reflectivity and loyalty to the data.
In current blind deconvolution methods 
ad-hoc parameter choices are often used, but the optimal regularization parameters should be determined by methods such as L-curve or general cross-validation (GCV) which actually require multiple realizations of the numerical experiments. 

Tests of the proposed algorithm with both synthetic and real data illustrate that this new approach gives fast (in terms of computing time) and high quality deconvolution results (in terms of a quality metric). 
In order to show the effectiveness of the proposed method,  recently proposed methods, such as, SMBD and F-SMBD are used as reference techniques.

\section{Proposed Blind Deconvolution Algorithm}
\begin{figure}[htbp]
        \centering
        \subfloat[]
        {
                \begin{minipage}{0.3\linewidth}
                        \centering
                        \includegraphics[width=\textwidth]{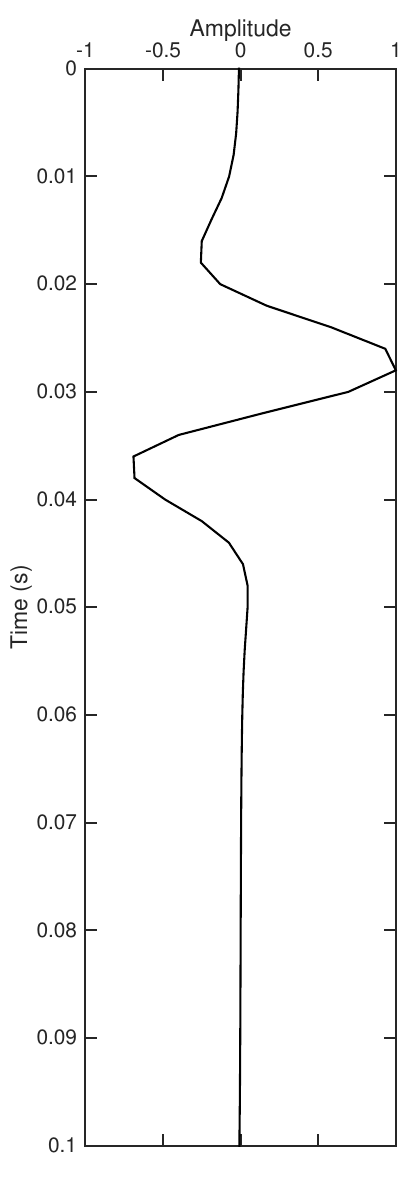}
                \end{minipage}
                \label{fig:refec_model_A}
        }
        \subfloat[]
        {    
                \begin{minipage}{0.3\linewidth}
                        \centering 
                        \includegraphics[width=\textwidth]{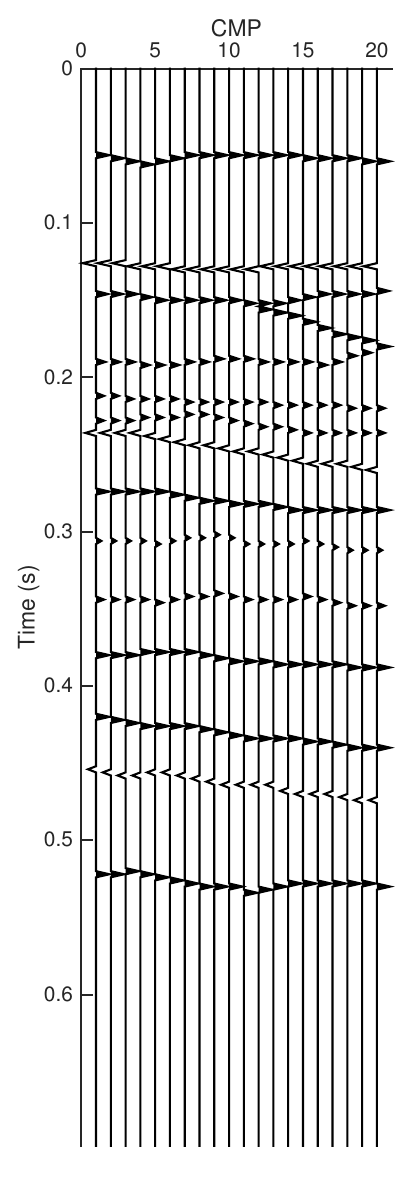}
                \end{minipage}
                \label{fig:refec_model_B}
        }
        \subfloat[]
        {
                \begin{minipage}{0.3\linewidth}
                        \centering 
                        \includegraphics[width=\textwidth]{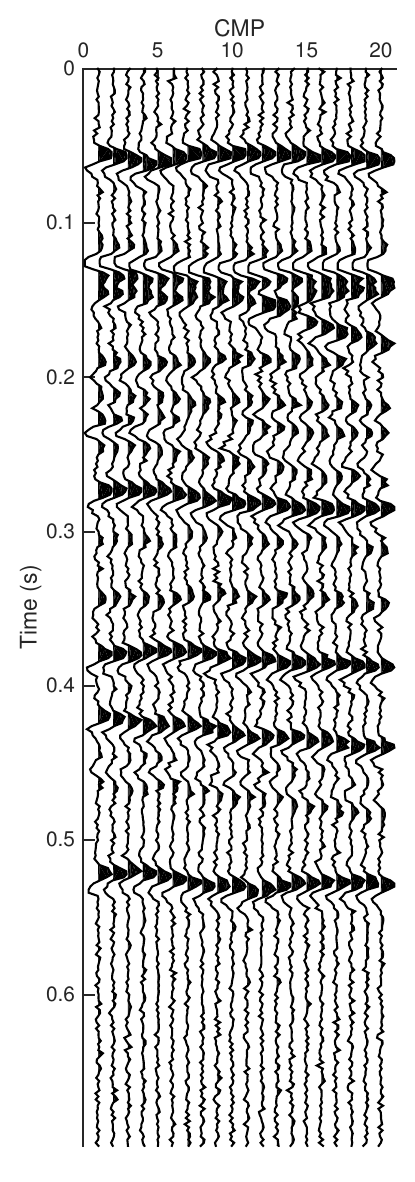}
                \end{minipage} 
                \label{fig:refec_model_C}
        }
        \caption{Convolutional model, (a) short-duration wavelet, (b) true reflectivity, (c) received traces with additive noise (SNR = 10\,dB).   
        } 
        \label{fig:refec_model}
\end{figure}

The classic model for a recorded seismic trace \cite{RobTre1980} is the output of a linear system where a seismic source wavelet is convolved with the 
earth's impulse response that is of finite duration.   In particular, the $j$-th received seismic trace is written as  
\begin{align}
\nonumber d_j [n] & = w[n]\ast r_j[n]+z_j[n]
= \sum_k w[n-k]r_j[k]+z_j[n], \\ &\mbox{\quad for\ }  n=1,\ldots,N,
\label{refmodel}
\end{align}
where $N$ is the number of received data samples per trace. 
In matrix-vector form, \eqref{refmodel} becomes
\begin{equation}\label{conv_model_vec} 
\bd_j= \bW \br_j + \bz_j, \mbox{\quad for }  j=1,\ldots,J,
\end{equation}
where $J$ is the number of channels, 
$\bW$ is an $N\!\times\!N$ convolution matrix formed from $w[n]$, 
$\br_j$ the vector of reflectivities for the $j$-th channel, $\bd_j$ the received data vector, and $\bz_j$ a noise vector.
Elements of the convolutional model are depicted in Figure \ref{fig:refec_model}. 
All $J$ channels can be combined into one matrix equation
\begin{equation}
\bd=
\begin{bmatrix}
\bd_1\\
\bd_2\\
\bd_3\\
\vdots\\
\bd_J
\end{bmatrix}
=
\begin{bmatrix}
\bW & \bzero & \bzero  & \cdots & \bzero\\
\bzero & \bW & \bzero  & \cdots & \bzero\\
\bzero & \bzero & \bW  & \cdots & \bzero\\
\vdots & \ddots & \ddots  & \ddots & \vdots\\
\bzero & \cdots & \bzero & \bzero  & \bW\\
\end{bmatrix}
\begin{bmatrix}
\br_1\\
\br_2\\
\br_3\\
\vdots\\
\br_J
\end{bmatrix}
= \mathbf{\Omega}\bx
\label{conv_model_MatrixVec}
\end{equation}
where the received data vectors for all $J$ channels are concatenated into one $JN\times1$ vector $\bd$, the reflectivity series into one $JN\times1$ vector $\bx = [\br_1^T,\, \br_2^T ,\,\ldots,\, \br_J^T]^T$,  and $\mathbf{\Omega}$ is a block diagonal matrix with the convolution matrix $\bW$ repeated along its diagonal.

The proposed method alternates between two steps: wavelet estimation assuming the $J$ reflectivities are known, and reflectivity estimation given a known common wavelet for all the channels. 
From the experiments it is noted that convergence of both estimates within a few iterations is usually obtained.
Similar alternating strategies are discussed in Section \ref{sec:EuclidMethods}.
The specific method presented here is named sparse multichannel blind deconvolution via spectral projected-gradient (SMBD-SPG); it is implemented as in Algorithm \ref{alg:alg1}.
\begin{algorithm}
        \caption{Sparse blind deconvolution by basis pursuit denoising}
        \label{alg:alg1}
        \SetAlgoLined
        \SetAlFnt{\small}
        \SetKwInOut{Kinit}{Initialization}
        \AlFnt
        \KwIn{ Seismic traces $\bd_j$, for $j=1,\ldots,J$ 
        }
        \Kinit{ Normalize the seismic section according to its global maximal value
                \begin{equation}
                \bd_j \leftarrow \bd_j/\max(\text{abs}(\bd)) \quad \mbox{for } j=1,\ldots,J.
                \end{equation}}
        Initialize the estimated reflectivity ${\br}^{(0)}_j$ for each $j\in[1,J]$ using a local peak finder
        
        { \For {$k \gets 1$ \KwTo $K$} 
                {Estimate common wavelet in Fourier domain ($\bwFT$ denotes FFT) 
                        \begin{equation} 
                        \bwFT^{(k)} = (\bRFT^H\bRFT + \lambda \bI)^{-1}\bRFT^H {\bdFT};
                        \end{equation} 
                        
                        
                        Smooth $\bwFT_s^{(k)}$ using moving average filter and $ \bw^{(k)}$, $\Re\{ \text{ifft}(\bwFT_s^{(k)})\}$, is multiplied by time window (further smooths $\bwFT^{(k)})$;\\[2mm]
                        
                        Update all $J$ reflectivity series 
                        $\bx = [\br_1^T,\, \br_2^T ,\,\ldots,\, \br_J^T]^T$ at once,
                        using SPG to solve:
                        \begin{equation}\label{obj5}
                        \bx^{(k)} = \argmin_\bx \|\bx\|_1 \quad\mbox{ subject to }
                        \|\bd-\bOmega^{(k)} \bx \|_2 \le \|\bz\|_2,
                        \end{equation}
                        where 
                        $\bOmega^{(k)}=\bdiag\{\bW^{(k)}\}$, and $\bW^{(k)}$ is the convolution matrix of ${\bw}^{(k)}$;
                        
                } 
        }
        \KwOut{ Estimated source wavelet $\bw^{(K)}$ and the $J$ reflectivity series $\br_j^{(K)}$ for $j=1,\ldots,J$. }
\end{algorithm}

\begin{figure}[htbp]
        \centering
        \includegraphics[width=0.5\textwidth]{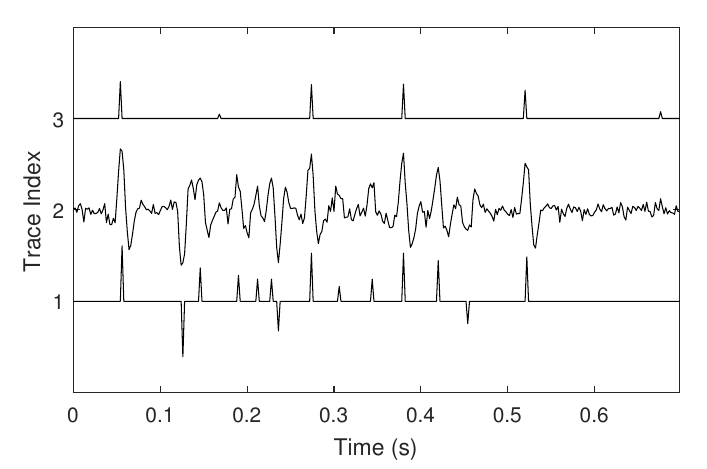}
        \caption{Initial estimated reflectivity using \texttt{findpeaks} in Matlab (index = 3) from the original trace (index = 2). The true reflectivity (index = 1) is shown as a reference.}
        \label{fig:ref_init}
\end{figure}

To initialize the iteration, we need estimates of the reflectivities ${\br}^{(0)}_j$ for all the traces, so we apply a simple peak locater (e.g., \texttt{findpeaks} in Matlab) on each received trace $\bd_j$ to obtain an initial estimated reflectivity as shown in Figure \ref{fig:ref_init}. In order to avoid picking multiple local peaks within the source wavelet itself, we constrain the distance between adjacent peaks to be greater than the wavelet duration. 
In practice, any efficient peak detector can be employed here to estimate $\br^{(0)}_j$. This initial step only needs to provide a rough estimate of the reflectivity model which is then refined in the subsequent iterative procedure. Although a better initial estimate might lead to faster convergence, the deconvolved output is not sensitive to this step once we run the algorithm for a few iterations.

The common wavelet for all channels is estimated using the reflectivity estimates from the previous iteration.
A minimum-energy wavelet is found subject to matching the data as in the convolution model \eqref{refmodel}. 
In the frequency domain, the seismic trace is a product of Fourier transforms of the wavelet and reflectivity series
\begin{equation}
D_j(e^{j\omega}) = R_j(e^{j\omega}) W(e^{j\omega}) + Z_j(e^{j\omega})
\end{equation}
which is then sampled with an FFT, $\mathcal F\{\cdot\}$, to create length-$N$ vectors $\bdFT_j=\mathcal F(\bd_j)$, $\brFT_j=\mathcal F(\br_j)$, and $\bwFT=\mathcal F(\bw)$.
Taking all $J$ channels together, the FFTs of the data $\bd_j$ and reflectivity estimates $\br^{(k)}_j$ at the $k$-th iteration are used to form $\bdFT$ and $\bRFT^{(k)}$ as
 \begin{equation}
 \bdFT = \begin{bmatrix}
 \mathcal F(\bd_1)\\
 \vdots\\
 \mathcal F(\bd_J)
 \end{bmatrix} \quad \mbox{and}\quad
 \bRFT^{(k)} = \begin{bmatrix}
 \diag(\mathcal F(\br^{(k)}_1))\\
 \vdots\\
 \diag(\mathcal F(\br^{(k)}_J))\\
 \end{bmatrix}
 \label{eq:JvectorMatrixDefs}
 \end{equation}
 where the operator $\mathcal F$ is a length-$N$ FFT.

Using the FFTs of the previous estimates of reflectivity in all channels, $\brFT_j^{(k-1)}=\mathcal F\{\br_j^{(k-1)}\}$, the matrix $ {\bRFT}^{(k-1)}$ is formed as in \eqref{eq:JvectorMatrixDefs}
and the following frequency-domain problem is solved for the minimum-energy wavelet common to all the channels 
\begin{equation}\label{discrep}
{\bwFT}^{(k)} = \argmin_{\bwFT} \|\bwFT\|_2 \mbox{\qquad subject to } \|{\bdFT}- {\bRFT}^{(k-1)}{\bwFT}\|_2 \le \|\bz\|_2,
\end{equation}
where $\bz=[\bz^T_1,\ldots,\bz^T_J]^T$ is the concatenated noise vector. 
 Next,  \eqref{discrep} can be written into a Tikhonov regularization form 
        \begin{equation}\label{obj4Moved}
                {\bwFT}^{(k)} =\argmin_{\bwFT} \|\bdFT - \bRFT^{(k-1)} \bwFT\|^2_2 + \lambda \|\bwFT\|^2_2;
        \end{equation}
 for a certain $\lambda$, which leads to a closed-form solution
\begin{equation}
\bwFT^{(k)} = (\bRFT^H\bRFT + \lambda \bI)^{-1}\bRFT^H {\bdFT},
\label{eq:solutionTikhonov}
\end{equation} 
where $\bRFT=\bRFT^{(k-1)}$ for conciseness.
The  diagonal-blocks structure of $\bRFT$ implies that  $\bRFT^H\bRFT$ is diagonal, which eliminates the need to compute a matrix inverse in \eqref{eq:solutionTikhonov}. Tikhonov regularization is used here, which is the most commonly used method to solve ill-posed inverse problems \cite{Golub1999}. The reflectivity is a nature of the earth and is non-zero. However, in the proposed method the initial estimate is obtained using peak detector, which may make the problem ill-posed. Furthermore, deconvolution is an inverse problem, however, the noise term makes it become ill-posed \cite{Deng2014}.

In practice, a seismic source typically has a smooth band-limited spectrum, 
but adding a regularization term to \eqref{obj4Moved} to control spectral smoothness is problematic.
Instead, we apply a frequency-domain smoothing filter to the solution $\bwFT^{(k)}$ from \eqref{eq:solutionTikhonov}, i.e., zero-phase moving average filter is applied on the spectrum. A zero phase moving average filter is an  FIR filter of odd length with coefficients given as

\begin{equation*}
f(n) = \begin{cases}
{1\over M} &-(M-1)/2\leq n\leq (M-1)/2\\
0 &\text{otherwise}
\end{cases}
\end{equation*}
where $M$ is the (odd) filter length. 
Further, smoothing is achieved by using window in the time domain (shown later in Figures  \ref{fig:TD_FD}d), which is consistent with an assumed limit on the time duration of the common wavelet.
An IFFT yields the wavelet estimate $\bw^{(k)}$.

Given the estimated wavelet $\bw^{(k)}$, the updated reflectivity $\br^{(k)}_j$ is obtained by deconvolving the wavelet.
To obtain a sparse result for the reflectivity sequences, we solve the following basis pursuit denoising (BPDN) problem \emph{for all channels} at once:
\begin{equation}\label{cbyc}
        \bx^{(k)} = \argmin_\bx \|\bx\|_1 \qquad\mbox{ subject to } \|\bd-\bOmega^{(k)} \bx \|_2 \le \|\bz\|_2, 
\end{equation}
where 
$\bOmega^{(k)} = \bI_J \otimes \bW^{(k)}$ is a block diagonal matrix in which each block is the convolution matrix $\bW^{(k)}$ formed from $\bw^{(k)}$, and the vector $\bx$ is the concatenation of all $J$ reflectivity series  $\bx = [\br_1^T,\, \br_2^T ,\,\ldots,\, \br_J^T]^T$.
For fast numerical implementation, instead of generating the large diagonal matrix with $J$ copies of $\bW^{(k)}$ and computing the matrix multiplication directly, it is more efficient to calculate the individual convolutions $\bW^{(k)} \br_j^{(k)}$.

 As an aside, we note that including the Euclid deconvolution term in this problem, which will be discussed in Section \ref{sec:EuclidMethods}, would destroy the computational simplicity of the algorithm that solves BPDN.
        
Compared with the existing schemes such as SMBD, F-SMBD, and modified SMBD \cite{KazGhoAO2016}, the primary motivation of our proposed scheme is to provide a much faster numerical approach with a simple way to estimate any parameters that control the algorithm.


In \eqref{cbyc} the vector $\bz$ is the noise and we need an estimate of the noise level $\|\bz\|_2$ to form the constraint.
In seismic deconvolution, the reflectivity series of two adjacent channels are typically similar.  Using this feature, we can easily estimate the noise level for each channel using its neighboring channels, which serves as a good noise error bound $(\sigma = \|\bz\|_2)$ in the BPDN problem. In addition, this noise level estimate can be an average of several channels, which reinforces the  lateral continuity in the data. 

There are two factors that come into play when choosing a method for solving the BPDN problem. One is computational efficiency, the other is how to handle the constraint which often involves regularization parameters.
The BPDN problem as stated in \eqref{cbyc}, or \eqref{bpdn} in the Appendix, has a constraint that is physically meaningful but not well suited to an algorithm.
The SPGL1 algorithm solves \eqref{cbyc} with a sequence of LASSO problems \eqref{lasso} where the $\ell_1$ constraint for LASSO is related to the noise constraint in \eqref{cbyc}.
The convergence rate of this approach is superlinear, and we have observed that 5--6 iterations are sufficient for multichannel blind deconvolution. 
One motivation of our proposed scheme is to provide a very efficient numerical approach with a simple way to estimate any parameters that control the algorithm\,--\,a topic that is treated in the next section.

\section{Parameter choices}
In SMBD-SPG several parameters require attention. First, in a least squares problem $\bH\bx +\bn=\bb$, where  $\bn$ is a noise vector and $\bb$ the vector of observations, the solution to the Tikhonov regularization problem 
\begin{equation}
\min_\bx \|\bH\bx-\bb\|_2^2+\lambda \|\bx\|_2^2
\end{equation}
has a closed-form
\begin{equation}
\bx_\lambda=(\bH^H\bH +\lambda\bI)^{-1}\bH^H\bb.
\end{equation}
If $\|\bn\|_2 \le \delta$ and $\bH^H \bb\in \text{Range}(\bH^H\bH)$, we have a bound
\begin{equation}
\|\bx_\lambda -\bH^H\bb\|_2 \le \delta/\sqrt{\lambda} + O(\lambda),
\end{equation}
so $\lambda=C \delta^{2/3}$ is a asymptotic ``brick wall" for Tikhonov regularization \cite{Gro2010}. For wavelet estimation in \eqref{obj4Moved}, if the frequency-domain noise level $\|\bn\|_2$ can be estimated, we can pick $\lambda = C (\|\bn\|_2)^{2/3}$ which is optimal in the asymptotic sense. 

Next, the noise levels $\|\bz\|_2 = (\sum_j \|\bz_j\|^2_2)^{1/2}$ must be chosen for the constraint in BPDN \eqref{obj5}. Here we need estimates of the noise energy on the traces, for $j=1,\ldots,J$. If we assume the noise is spatially stationary then the noise energy has roughly the same level on neighboring channels. 
Because the reflectivity is determined by the real earth, as well as the relative locations of the sensors and the seismic source, a high degree of resemblance for the reflectivity in spatially close channels is commonly observed \cite{AkiRic2009}. 
Under these assumptions, we can calculate the variance of the difference of two adjacent traces to estimate the incoherent noise energy
\begin{equation}
\begin{aligned}
\var (\bd_1\hspace{-.12cm}-\hspace{-.12cm}\bd_2)&\hspace{-.12cm} =\hspace{-.12cm} \var(r_1[n]\hspace{-.05cm}\ast \hspace{-.05cm}w[n] \hspace{-.12cm}+\hspace{-.12cm} z_1[n]\hspace{-.12cm}-\hspace{-.12cm}r_2[n]\ast w[n] + z_2[n])\\
& \approx \var(z_1[n]-z_2[n]) = 2\var(z[n])
\end{aligned} 
\label{eq:var}  
\end{equation}
which then leads to  $\|\bz_j\|_2 \approx \sqrt{N \var(z_j[n])}$. For the high SNR case, where a silent segment (noise only part) of traces can be easily identified, we can use the silent segments to replace the whole traces $\bd_1$ and $\bd_2$ in \eqref{eq:var} for better estimation.
This time-domain estimate of the noise level can be used as the frequency-domain noise estimate by virtue of Parseval's Theorem.

Finally, we note that the wavelet length could be estimated based on knowledge of the seismic source used in the survey. 
However, in the examples presented here, we assume a wavelet length of 51 sample points.

In the sequel, methods based on Euclid deconvolution are briefly presented for comparison together with motivation for the proposed method.  

\section{Methods Based on Euclid Deconvolution}
\label{sec:EuclidMethods}

The algebraic structure of the multichannel deconvolution problem can be exploited to eliminate the wavelet and write the blind deconvolution problem as solving a large set of linear equations directly for the reflectivity series.
This was done in 1997 for seismic deconvolution by Rietsch  \cite{Rie1997} who coined the term ``Euclid technique.''
The same equations were also studied earlier in signal processing for applications such as speech dereverberation and channel equalization \cite{XuLiuAO1995,GurNikias1995}.
Recently, several authors have shown that a sparse solution of the wavelet-free linear equations is feasible which leads to a method called sparse multichannel blind deconvolution (SMBD) and its variants, F-SMBD and modified SMBD.

The primary motivation of this work is to find an efficient numerical scheme to replace the Euclid deconvolution (later in equation \eqref{eq:EuclidDeconv}) which is based on the identical wavelet assumption across all channels. It turns out by applying an iterative scheme which contains a simultaneous wavelet estimation across all channels, we can  relieve us from using the Euclid deconvolution and handling the huge matrix $\bA$, but still get high quality deconvolution results for both synthetic and filed seismic data sets. Recently, several iterative algorithms are proposed for the seismic deconvolution problem \cite{KazGhoAO2016,MirCoh2017}. Our proposed method improves those schemes in different aspects. For example, compared with the multichannel semi-blind deconvolution scheme in \cite{MirCoh2017}, which begins with an assumed source wavelet, we begin with an initial guess of the reflectivity which removes the assumption on the source wavelet without sacrificing on the quality of the deconvolution results. In addition, applying spectral projected-gradient scheme and frequency-domain computation achieve a significant speed-up comparing with iterative gradient descent.

\subsection{Euclid Deconvolution}
The $z$-transform of equation \eqref{refmodel}  gives  
\begin{equation}
D_j(z) = W(z) R_j(z)+Z_{j}(z).
\label{eq:zTransXWR}
\end{equation}
By considering \eqref{eq:zTransXWR} for a pair of traces, we can eliminate  the wavelet term $W(z)$ and obtain the following system of equations
\begin{align}
\nonumber D_p(z)R_q(z) - D_q(z)R_p(z) &= Z_p(z)R_q(z) - Z_q(z)R_p(z) , \\ &\quad \text{for } p\neq q.
\label{eq:EuclidDeconv}
\end{align}
It is convenient to write equation \eqref{eq:EuclidDeconv} in matrix form:
\begin{equation}
\bD_p \br_q- \bD_q \br_p =  \bZ_p \br_q- \bZ_q \br_p , 
\label{eq:matrixXR}
\end{equation}
where $\bD_{p(q)}$ and $\bZ_{p(q)}$ are $N\!\times\!N$ convolution matrices formed from the received data and the noise in channels $p$ and $q$. 
Combining all instances of \eqref{eq:matrixXR} into one equation, we have
\begin{equation}\label{Ar=e}
\bA \bx = \be, 
\end{equation}
where $\bx$ is a $JN$-element vector of concatenated reflectivity series for all $J$ channels
\begin{equation}\label{eq:xIsConcatenatedRefls}
\bx = [\br_1^T,\, \br_2^T ,\,\ldots,\, \br_J^T]^T,
\end{equation}
%
$\bA$ is a $\frac{1}{2}(J-1)JN\times JN$ matrix consisting of convolution matrices in blocks, and
$\be$ is a $\frac{1}{2}(J-1)JN$-element vector formed by concatenating all the right-hand side vectors in \eqref{eq:matrixXR}.

If the convolution model is a perfect fit to the data, then the noise terms $\bz_j$ in equation \eqref{conv_model_vec} are zero and $\|\be\|=0$.
Conversely, if $\|\be\|=0$ and all $\|\br_j\|\neq0$, then all the noise terms $\bz_j$ in equation \eqref{conv_model_vec} must be zero, which implies that the convolution model is a perfect fit to the data.
These facts motivate an optimization problem that minimizes $\|\bA \bx \|_2$.

\subsection{Numerical Methods: SMBD and F-SMBD}

In this section, we revisit the SMBD and F-SMBD methods which will serve as benchmarks for the comparison with our proposed method. 
In general, the vector $\be$ in \eqref{Ar=e} is an error to be minimized \cite{KazSac2014}.
In SMBD, the energy $\|\be\|_2^2$ is minimized
while observing a regularization constraint, i.e., 
\begin{equation}\label{obj1}
\hat{\bx} = \argmin_{\bx}\left\{ \frac{1}{2} \|\bA \bx\|_2^2 +\lambda \cR_\epsilon (\bx)\right\}, \mbox{\quad subject to } \bx^T\bx =1.
\end{equation}
The constraint $\bx^T\bx =1$ rules out a trivial solution in equation \eqref{obj1}. 
To make the optimization easier, the regularization term is defined with a differentiable smoothed $\ell_1$-like norm
\begin{equation}\label{l1l2norm}
\cR_\epsilon (\bx) = \sum_n \left(\sqrt{(x[n])^2+\epsilon^2}-\epsilon \right)
\end{equation}
which is used to promote sparsity of the output $\hat{\bx}$. Small values of the parameter $\epsilon$ generate a mixed norm that behaves like the $\ell_1$ norm when $|x[n]|>\epsilon$, i.e., $\cR_\epsilon (\bx) \approx  \| \bx \|_1^{\phantom{y}}$.

Recently, a modified SMBD has been proposed \cite{KazGhoAO2016} in which the $J$ reflectivity series and the source wavelet are estimated via
\begin{equation}\label{eq:modSMBDargmin}
\{\hat\bx,\hat{\bw}\} \hspace{-.12cm}= \hspace{-.1cm}\argmin_{\bx,\bw} \left\{ \|\bA \bx\|_2^2 + \lambda_x \|\bx\|_1  \hspace{-.12cm}+ \hspace{-.12cm}\lambda_n \|\bOmega\bx\hspace{-.12cm}-\hspace{-.12cm}\bd\|_2^2  \hspace{-.12cm}+ \hspace{-.12cm}\lambda_w \|\bw\|_2^2 \right\}\hspace{-.12cm},
\end{equation}
which is solved by an alternating minimization technique. By fixing
the source wavelet the problem can be solved for the reflectivity using any $\ell_2$-$\ell_1$ solvers. By fixing
the reflectivity, using the updated version of it, the estimation of the source wavelet can be cast as an
$\ell_2$-$\ell_2$ problem which has a closed form solution. This alternating process will be halted when it converges.
There are three regularization parameters one needs to choose in this scheme.

In existing methods for sparse blind deconvolution, a pseudo $\ell_1$ norm regularization of the reflectivity, as in \eqref{l1l2norm}, is used to obtain a least-squares problem.
However, with state-of-the-art algorithms it is feasible to attack the $\ell_1$ norm optimization directly, e.g. using SPGL1 as proposed in this paper.
Since the smoothed $\ell_1$ norm is differentiable, one can use steepest descent to minimize the objective function, but this process usually needs a large number of iterations to converge. Additionally, a good choice of the parameter $\lambda$ in \eqref{obj1} can only be determined by L-curve or GCV methods, which require multiple realizations as well.

The F-SMBD method devises a single deconvolution filter $v[n]$ that operates on all the traces $y_j[n] = d_j[n] \ast v[n]=\sum_k v[k]d_j[n-k]$ by minimizing the sum of the smoothed $\ell_1$ norms of the deconvolved traces  $y_j[n]$.
Thus the following optimization problem is solved for the vector of $P$ filter coefficients
$\bv=[v[1],\ldots,v[P]]^T$,
\begin{equation}\label{obj2}
\hat{\bv}= \argmin_\bv \mathcal L(\bv)=\argmin_{\bv} \sum_j \cR_\epsilon (\by_j), \mbox{\ subject to }  \bv^T\bv =1,
\end{equation}
where the sparsity-promoting norm $\cR_\epsilon$ is defined as
\begin{equation}
\cR_\epsilon (\by_j) = \sum_n \left(\sqrt{\frac{y_j^2[n]}{\sigma^2_{\by_j}}+\epsilon^2}-\epsilon \right)
\label{eq:RepsFSMBD}
\end{equation}
and $\sigma^2_{\by_j}=\sum_n y^2_j[n]/N$ is the estimated variance of the deconvolved reflectivity series. As in SMBD the constraint $\bv^T\bv =1$ is imposed to avoid a trivial solution.
The deconvolved traces $y_j[n] = d_j[n] \ast v[n]$ are per channel estimates of the reflectivity series.
Steepest descent provides a reliable iterative solution to the minimization problems in \eqref{obj1} and \eqref{obj2}.\footnote{In equation (16) of \cite{NosTak2016} there is a typo for the $k$-th component of $\nabla{\mathcal{L}}$, which should be
       \begin{align}
     \nonumber  \nabla{\mathcal L}_k&=\sum_j \sum_n \frac{\{y^{2}_j[n]/\sigma_{\by_j}^2+\epsilon^2\}^{-1/2}}     {\sigma^4_{\by_j}}\\& ~
       \left\{\sigma^2_{\by_j}y_j[n]d_j[n-k] -  \frac{y_j^{2}[n]}{N}\sum_n  y_j[n]d_j[n-k]\right\}.\nonumber
       \end{align} }
%
%

The number of parameters in F-SMBD is $P$, the length of deconvolution filter, while the number in SMBD is $JN$, so the computational complexity of F-SMBD is significantly lower than SMBD.
Furthermore, SMBD and modified SMBD require that regularization parameter(s) 
in \eqref{obj1} or \eqref{eq:modSMBDargmin}  be chosen which is accomplished with the L-curve or cross validation strategy.
Finally, the matrix $\bA$ in \eqref{Ar=e} has $\frac{1}{2}(J-1)JN$ rows, so when the number of traces is large, the memory requirement for SMBD, or modified SMBD, might be too high to store $\bA$ for an entire seismic section. Although, there are some efficient approaches to compute the $\bA\bx$ and $\bA^H\bA$ either in the frequency domain, or by exploiting the sparsity of $\bA$, usually with hundreds of traces a patch-by-patch strategy would be employed to manage memory and computation.


\section{Synthetic data test}
For this test, 20 traces are generated with a sampling frequency of 500\,Hz using the reflectivity shown in Figure \ref{fig:refec_model}b, which can be downloaded from \cite{synModel2016}.
 The received data in Figure \ref{fig:refec_model}c is the convolution of the reflectivity with a Ricker wavelet of center frequency 40\,Hz with 50 degrees of phase shift (see Figure \ref{fig:refec_model}a) plus additive white Gaussian noise (AWGN)  of SNR = 10\,dB. The SNR adopted in this work for  the signal-plus-noise model, i.e., $\bx = \bs+\bn$ is defined as 
\begin{equation}
\text{SNR} = 10\log_{10}\left(\frac{\|\bs\|^2_2}{\|\bn\|_2^2}\right),
\end{equation}
where $\|\bs\|^2_2$ is the total signal energy and $\|\bn\|_2^2$ the noise energy.

For the F-SMBD algorithm, a length-51 deconvolution filter is used, which is initialized with a single spike located at the middle of the filter. Other parameters used for F-SMBD are $\epsilon=1$, and step size $\mu=0.02$. 
For the SMBD algorithm, the regularization parameter $\lambda$ is set to 4, the smoothed norm parameter $\epsilon$  is set to 0.0001, and $\alpha=0.2$. 
The number of iterations for SMBD and F-SMBD is set to 800 and 500, respectively, which ensures that both algorithms converge. 

\begin{figure*}[!htb]
        \centering
        \subfloat[]
        {
                \begin{minipage}{0.23\linewidth}
                        \centering
                        \includegraphics[width=\textwidth]{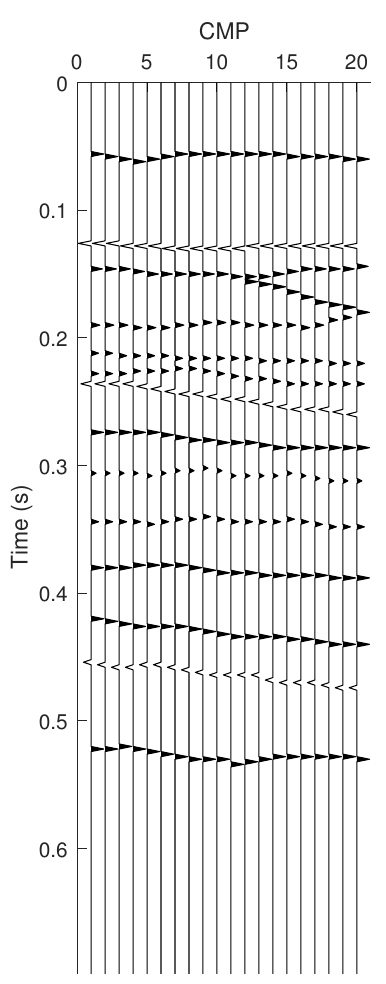}
                \end{minipage}
                \label{fig:syn_result_10dB_5iter_A}
        }
        \subfloat[]
        {
                \begin{minipage}{0.23\linewidth}
                        \centering 
                        \includegraphics[width=\textwidth]{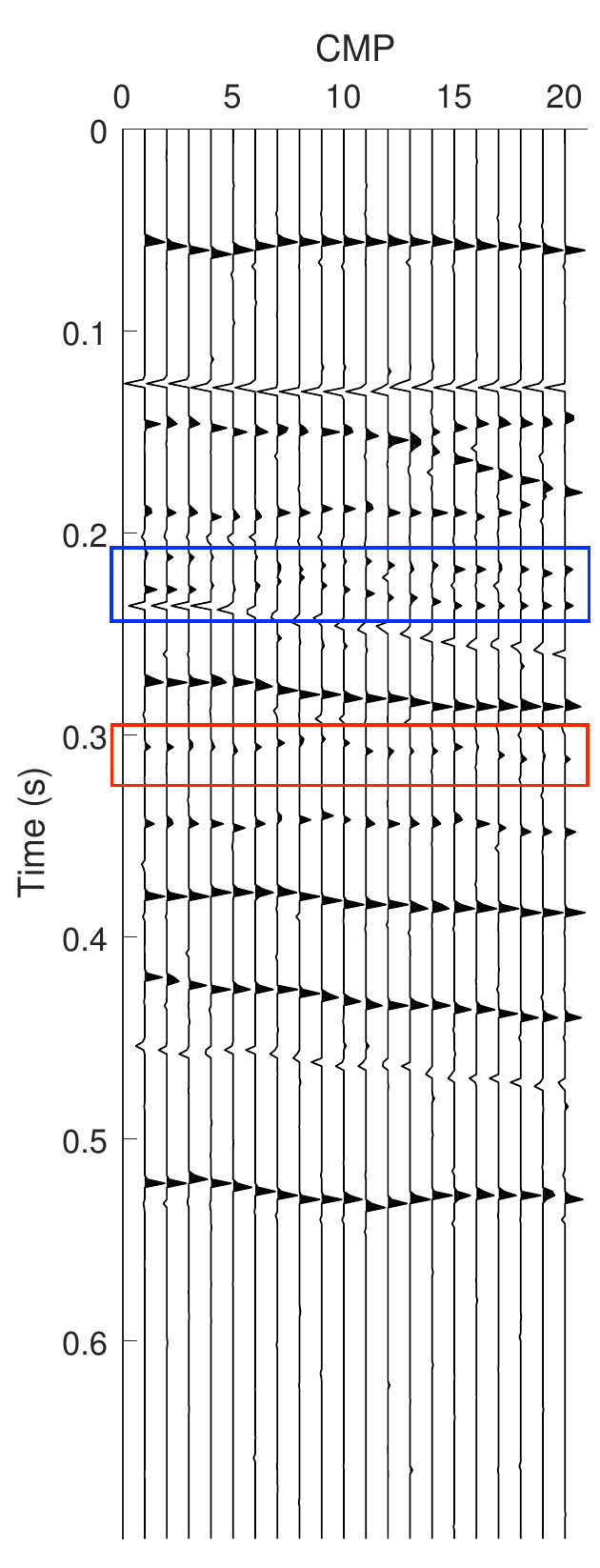}
                \end{minipage}
                \label{fig:syn_result_10dB_5iter_D}
        }
        \subfloat[]
        {    
                \begin{minipage}{0.23\linewidth}
                        \centering 
                        \includegraphics[width=\textwidth]{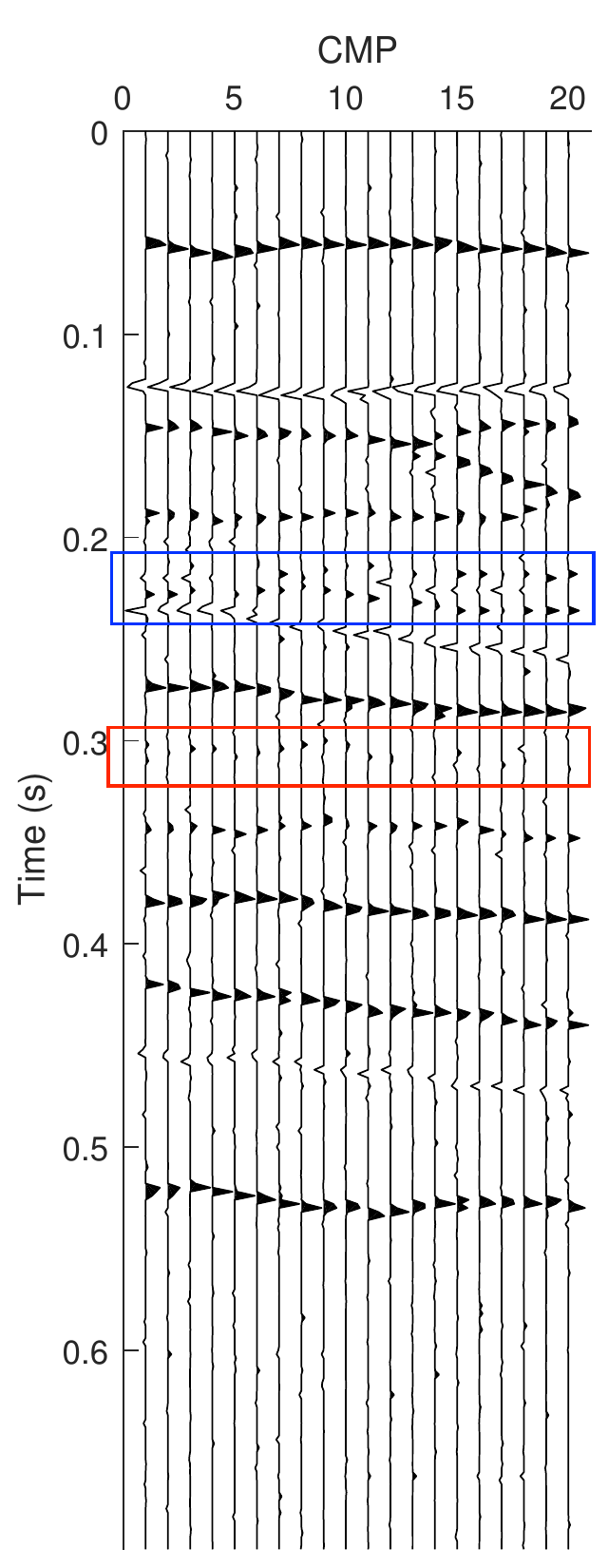}
                \end{minipage}
                \label{fig:syn_result_10dB_5iter_B}
        }
      \subfloat[]
      {
              \begin{minipage}{0.23\linewidth}
                      \centering 
                      \includegraphics[width=\textwidth]{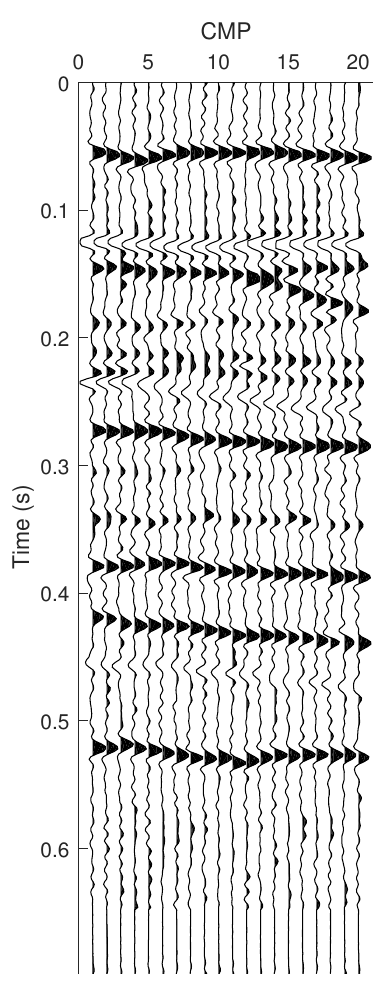}
              \end{minipage} 
              \label{fig:syn_result_10dB_5iter_C}
      }
        \caption{Deconvolution result for SNR = 10\,dB. (a) True reflectivity. Deconvolution result for (b) SMBD-SPG (5 iterations), (c) SMBD, (d) F-SMBD.
        } 
        \label{fig:syn1}
\end{figure*}

\begin{figure}[htbp]
        \begin{center}
                \includegraphics[width=0.5\textwidth]{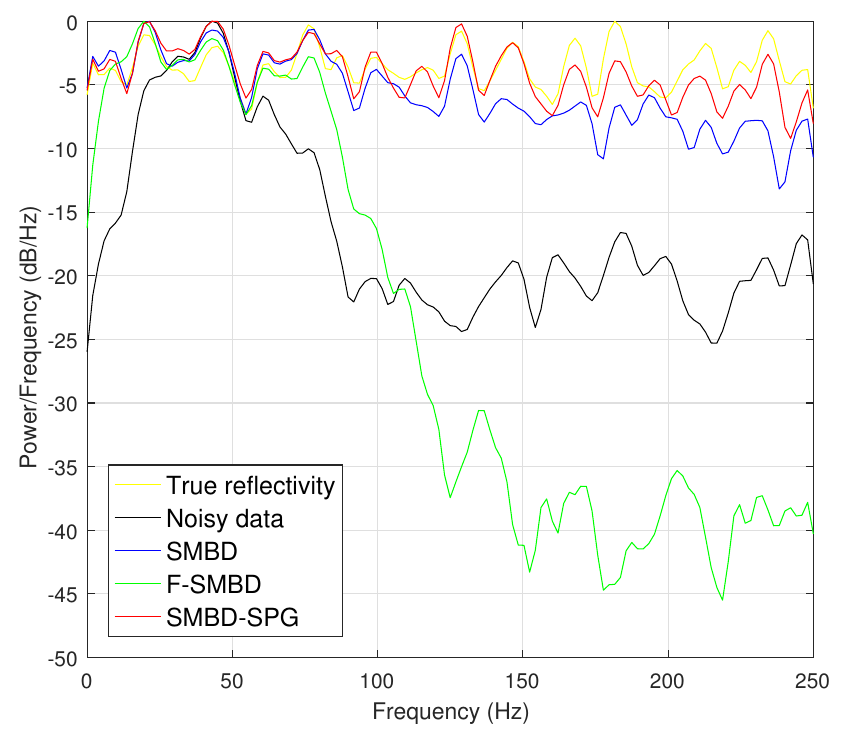}
                \caption{Smoothed normalized PSD of the true reflectivity, noisy data and recovered reflectivity series. SNR = 10\,dB, SMBD-SPG with 5 iterations. PSD computed per channel via MATLAB's \texttt{pwelch} function with section length = 100 and overlap = 60, and then averaged over all channels.}
                \label{fig:synpsd}
        \end{center}
\end{figure}

The recovered reflectivity series using the three methods are shown in Figures \ref{fig:syn1}b, \ref{fig:syn1}c, and \ref{fig:syn1}d. For SMBD-SPG the number of iterations is set  to 5, i.e., $K\!=\!5$ in Algorithm \ref{alg:alg1}. Within the blue and red rectangles, we notice that SMBD-SPG is  better than SMBD and F-SMBD for weak and close reflectors, in the sense that information with better precision is preserved for interpretation. 
Next, the quality of the three methods is evaluated using the normalized power spectrum density (PSD), shown in Figure \ref{fig:synpsd}. SMBD-SPG yields the flattest PSD among the three schemes, which is closest to the PSD of the true reflectivity.  SMBD is nearly as flat in the frequency domain, while F-SMBD exhibits an obvious band limit to frequencies below 80\,Hz.

\begin{figure}[htbp]
        \centering
        \subfloat[]
        {
                \begin{minipage}{0.477\linewidth}
                        \centering
                        \includegraphics[width=\textwidth]{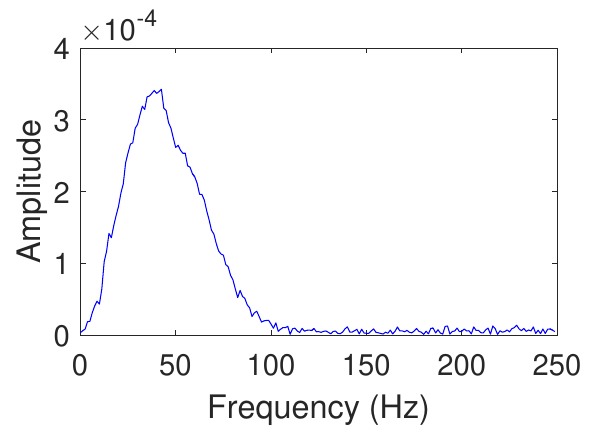}
                \end{minipage}
                \label{fig:TD_FD_iter5_A}
        }
        \subfloat[]
        {    
                \begin{minipage}{0.477\linewidth}
                        \centering 
                        \includegraphics[width=\textwidth]{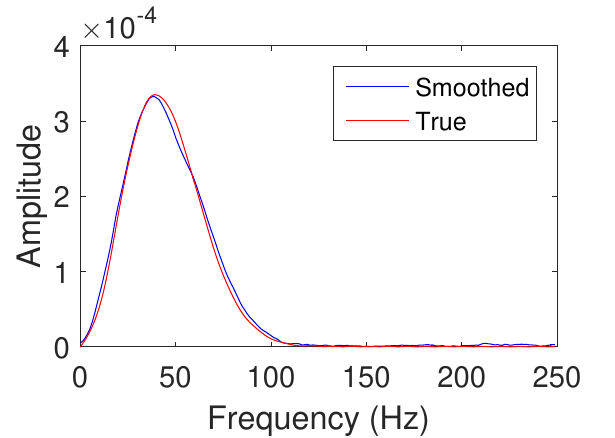}
                \end{minipage}
                \label{fig:TD_FD_iter5_B}
        }\\
        \subfloat[]
        {
                \begin{minipage}{0.477\linewidth}
                        \centering 
                        \includegraphics[width=\textwidth]{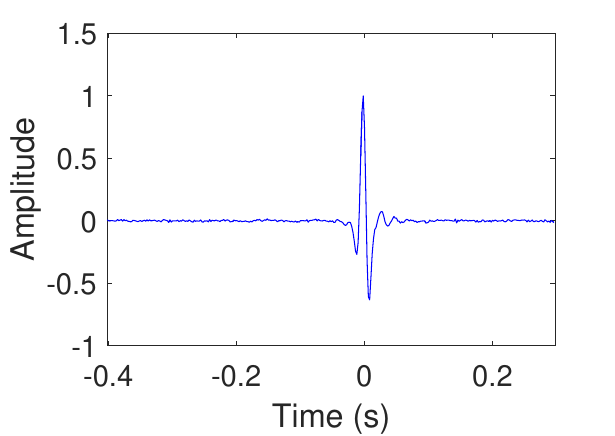}
                \end{minipage} 
                \label{fig:TD_FD_iter5_C}
        }
        \subfloat[]
        {
                \begin{minipage}{0.477\linewidth}
                        \centering 
                        \includegraphics[width=\textwidth]{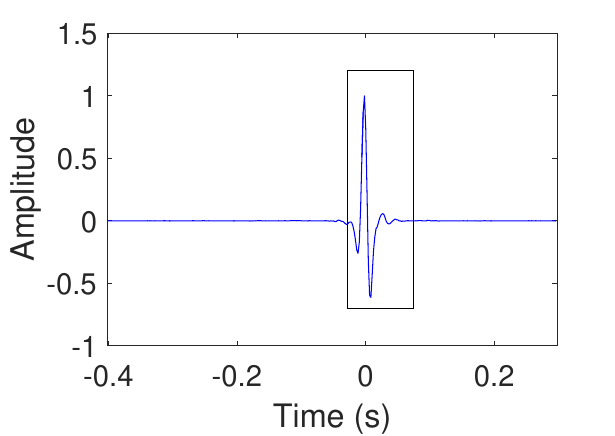}
                \end{minipage}
                \label{fig:TD_FD_iter5_D}
        }
        \caption{Wavelet estimation after 5 iterations. (a) spectrum before smoothing, (b) smoothed and true spectrum, (c) time-domain wavelet before smoothing, (d) wavelet after smoothing.  
        } 
        \label{fig:TD_FD}
\end{figure}

\begin{figure}[htbp]
        \centering
        \subfloat[]
        {
                \begin{minipage}{0.477\linewidth}
                        \centering
                        \includegraphics[width=\textwidth]{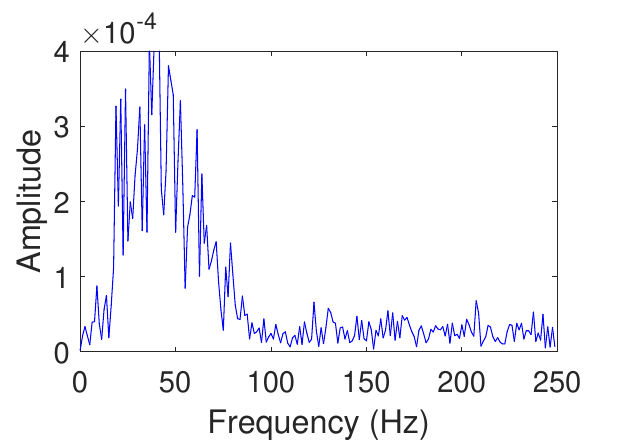}
                \end{minipage}
                \label{fig:TD_FD_iter1_A}
        }
        \subfloat[]
        {    
                \begin{minipage}{0.477\linewidth}
                        \centering 
                        \includegraphics[width=\textwidth]{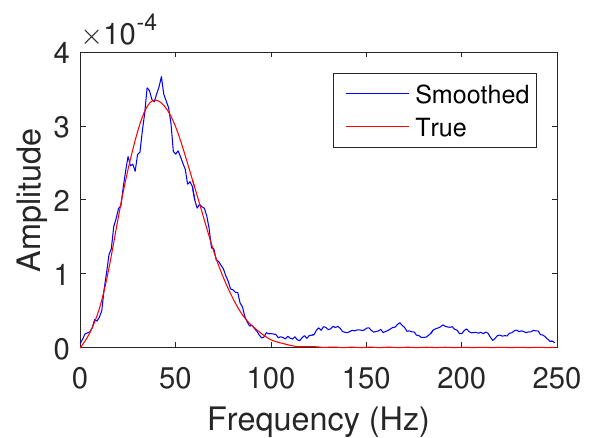}
                \end{minipage}
                \label{fig:TD_FD_iter1_B}
        }\\
        \subfloat[]
        {
                \begin{minipage}{0.477\linewidth}
                        \centering 
                        \includegraphics[width=\textwidth]{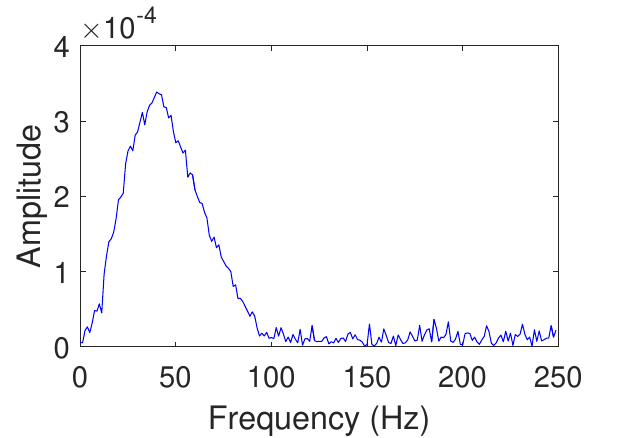}
                \end{minipage} 
                \label{fig:TD_FD_iter3_A}
        }
        \subfloat[]
        {
                \begin{minipage}{0.477\linewidth}
                        \centering 
                        \includegraphics[width=\textwidth]{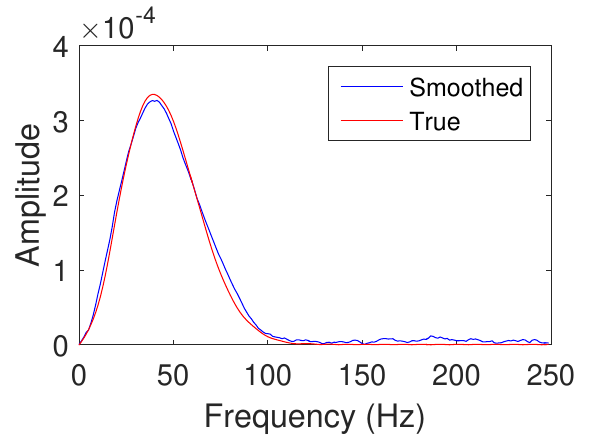}
                \end{minipage}
                \label{fig:TD_FD_iter3_B}
        }
        \caption{Wavelet estimation after one iteration: (a) spectrum before smoothing, (b) smoothed and true spectrum.  Wavelet estimation after three iterations: (c) spectrum before smoothing, (d) smoothed and true spectrum.   
        } 
        \label{fig:TD_FD_iter1_3}
\end{figure}

Although the SMBD-SPG scheme is iterative, in many cases only a few iterations are needed to get a good approximation of the wavelet due to the smoothing filter. 
For example, the result in Figure \ref{fig:syn1}b is obtained after only 5 iterations. The smoothing in frequency-domain is achieved by using a moving average filter on the spectrum.  The estimated wavelet with and without smoothing in frequency  can be found in Figure \ref{fig:TD_FD}. Note the accurate match in Figure \ref{fig:TD_FD}b after applying the frequency-domain smoothing filter, which is length-11 with uniform values. After getting the time-domain wavelet, a portion of it is taken (which consists of 51 samples and indicated as a rectangle in Figure \ref{fig:TD_FD}d) to be used for updating the reflectivity with BPDN \eqref{obj5}. 
In Figure \ref{fig:TD_FD_iter1_3} we show the estimated wavelet's spectrum with and without smoothing after 1 and 3 iterations, so we can observe the amount of improvement over iterations.

%
%

\begin{figure}[htb]
        \centering
        \subfloat[]
        {
                \begin{minipage}{0.46\linewidth}
                        \centering
                        \includegraphics[width=\textwidth]{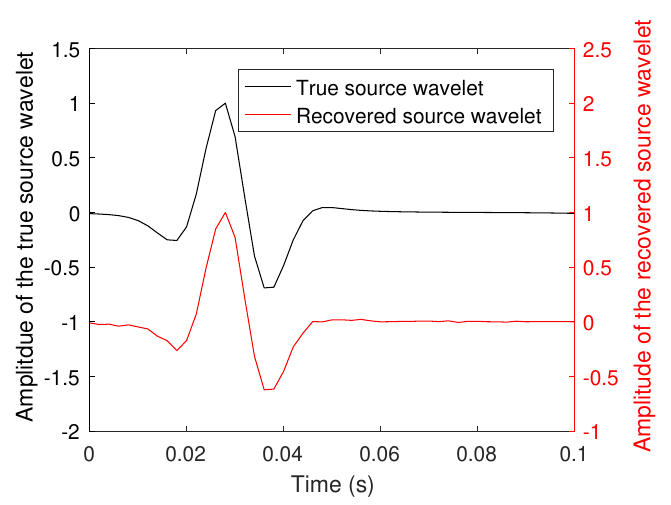}
                \end{minipage}
                \label{fig:rec_wave}
        }
        \subfloat[]
        {
                \begin{minipage}{0.45\linewidth}
                        \centering 
                        \includegraphics[width=\textwidth]{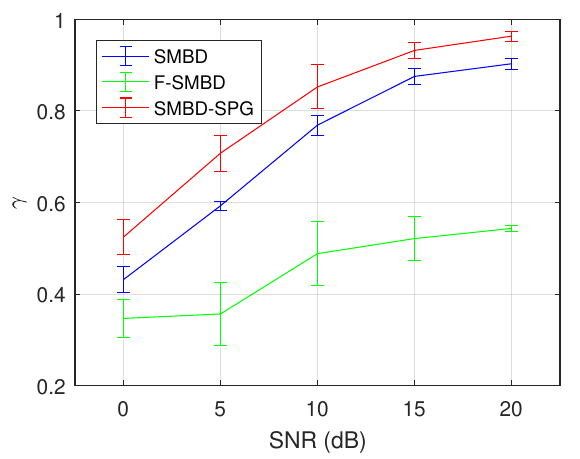}
                \end{minipage} 
                \label{fig:SNR_gamma}
        }
        
      \subfloat[]
      {
              \begin{minipage}{0.45\linewidth}
                      \centering 
                      \includegraphics[width=\textwidth]{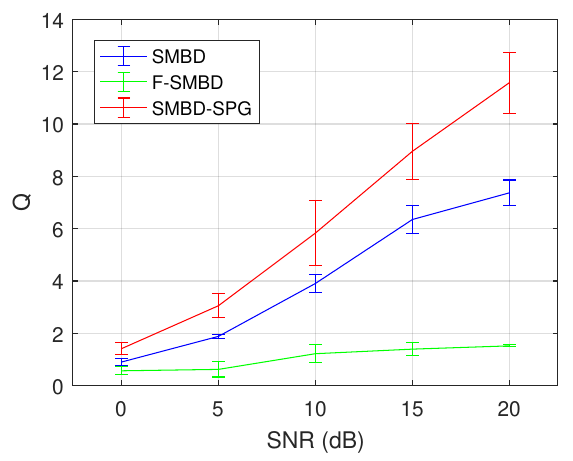}
              \end{minipage}
              \label{fig:SNR_Q}
      }
       \subfloat[]
        {    
                \begin{minipage}{0.45\linewidth}
                        \centering 
                        \includegraphics[width=\textwidth]{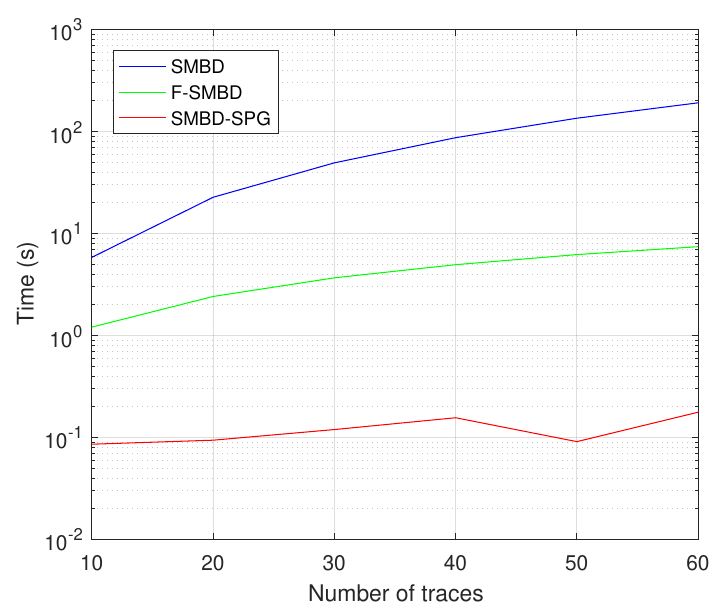}
                \end{minipage}
                \label{fig:time_NoTraces}
        }\\
        \caption{Comparison among SMBD, F-SMBD, and SMBD-SPG for different attributes of the synthetic data example. (a) Recovered wavelet (by SMBD-SPG) and true wavelet when SNR = 10\,dB; (b) Normalized correlation coefficient $\gamma$  vs.\ SNR (mean and standard deviation averaged over 20 traces); (c) Simulation time vs.\ number of traces with SNR = 10\,dB, each trace contains 350 samples; (d) Quality metric $Q$ vs.\ SNR (mean and standard deviation over 20 traces).
        } 
        \label{fig:mean and std of nets}
\end{figure}

To compare the performance of SMBD-SPG with SMBD and F-SMBD, the normalized correlation coefficient $\gamma$ and the quality $Q$ is used to measure the similarity between the reflectivity $\bx$ and its estimate $\hat\bx$.  
\begin{subequations}
        \begin{align}
        \gamma =& 
       \frac{\hat{\bx}^T \bx }{\|\hat\bx\|_2\|\bx\|_2}, \label{eq:defnGamma}
       \\
       Q 
       =& -20 \log_{10} \left( \frac{ \|{\bx}-{\hat\bx}{(\bx^T\bx)}/{( \hat{\bx}^T\hat{\bx})}\|_2}{\|{\bx}\|_2 }\right),
        \end{align}
\end{subequations}
where $\bx$ and $\hat{\bx}$ are long vectors formed by concatenating true and estimated reflectivity series, respectively. 
Recovered wavelet by SMBD-SPG) and true wavelet when SNR = 10\,dB is depicted in Figure \ref{fig:mean and std of nets}a. After running 10 Monte-Carlo realizations of the random noise for various levels, the mean value and standard deviation for $\gamma$ and $Q$ versus SNR are shown in Figure \ref{fig:mean and std of nets}b and \ref{fig:mean and std of nets}c.
The SMBD-SPG algorithm (after 5 iterations) outperforms the SMBD algorithm in terms of $\gamma$ for all noise levels. 

To show that the proposed algorithm is faster than SMBD and F-SMBD, measured calculation time for the algorithms is plotted against the number of traces (each trace contains 350 time samples) in
Figure \ref{fig:mean and std of nets}d. For 60 traces the computation time of 5 iterations of SMBD-SPG equals 0.1773\,s, 
which is 0.0926\% of the SMBD time and 2.38\% of F-SMBD.
Throughout this paper, all experiments were performed on Matlab R2016b with a 3.5\,GHz Intel i7 quad-core CPU and 32\,GB RAM. Table \ref{table_conv} summarizes the computational complexity of various deconvolution methods.

\section{Real data results}
In this section, results obtained using SMBD, F-SMBD and SMBD-SPG on a real seismic data set are presented.  The seismic data is from the National Petroleum Reserve, Alaska (NPRA) Legacy Data Archive by USGS (1976), Line ID 31-81 \cite{NPRA1976}.
For the real data scenario, we run SMBD  on blocks of duration 0.6\,s in time and 100 traces. For F-SMBD, the deconvolution filter length is taken as 51 samples with a spike at the middle of the filter for initialization. The learning rate for F-SMBD is set to $\mu=0.02$. The deconvolution filter for F-SMBD is obtained for part of the data, i.e., time range $[0.6,2.4]$, and traces 251 to 254. 
For SMBD-SPG, the blocks are the same as in SMBD. The remaining parameters of both algorithms are unchanged from the synthetic data test. The average processing times for each patch are 95.22\,s and 0.7618\,s for SMBD and SMBD-SPG, respectively. Recently, the authors of SMBD proposed a modified SMBD \cite{KazGhoAO2016} method by adopting an iterative scheme that alternates between wavelet estimation and reflectivity estimation. It turns out the performances have been greatly improved, however each iteration of the modified SMBD requires similar computational efforts to SMBD. 

\begin{figure}[htbp]
        \centering
        \subfloat[]
        {
                \begin{minipage}{0.45\linewidth}
                        \centering
                        \includegraphics[width=0.95\textwidth,height=1.8\textwidth]{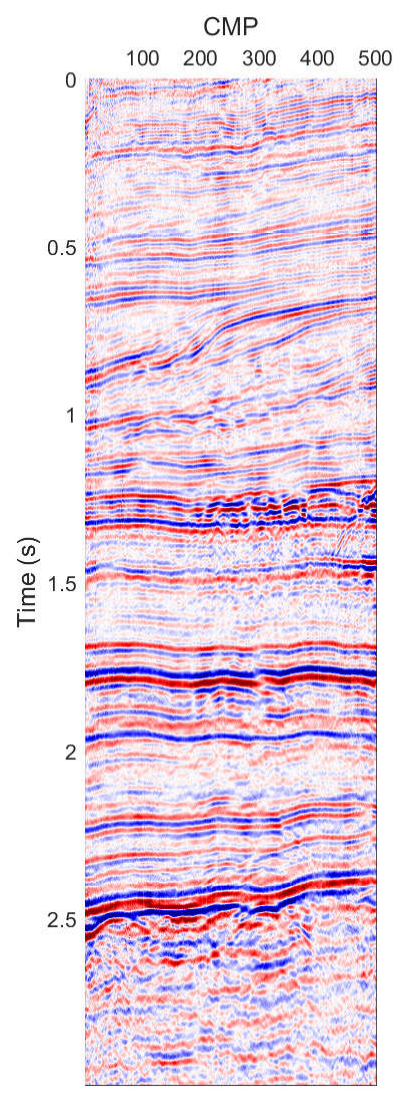}
                \end{minipage}
                \label{fig:real_full_A}
        }
        \subfloat[]
        {
                \begin{minipage}{0.45\linewidth}
                        \centering 
                        \includegraphics[width=0.95\textwidth,height=1.8\textwidth]{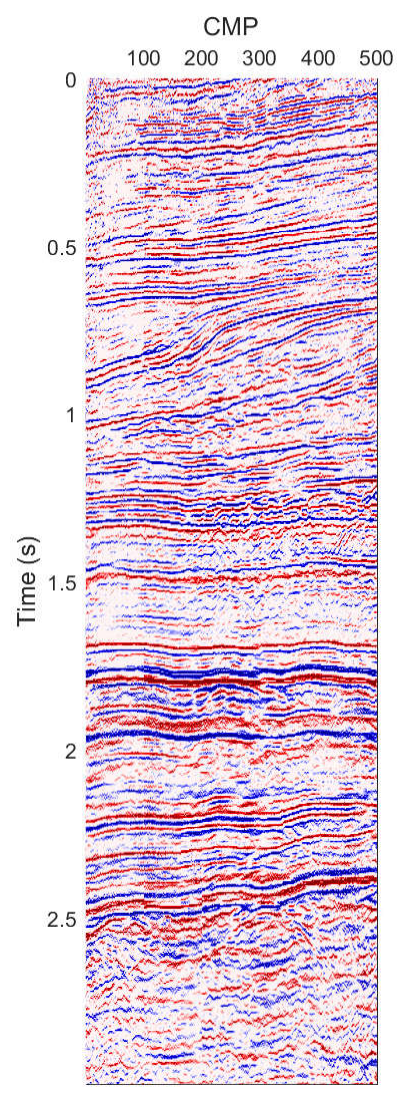}
                \end{minipage}
                \label{fig:real_full_D}
        }
        \caption{ (a) Stacked CMP data for 3\,s and 500 traces. Deconvolution results for (b) SMBD-SPG (5 iterations).     
        } 
        \label{fig:real1}
\end{figure}
\begin{figure*}[htbp]
        \centering
        \subfloat[]
        {
                \begin{minipage}{0.45\linewidth}
                        \centering
                        \includegraphics[width=0.95\textwidth,height=1\textwidth]{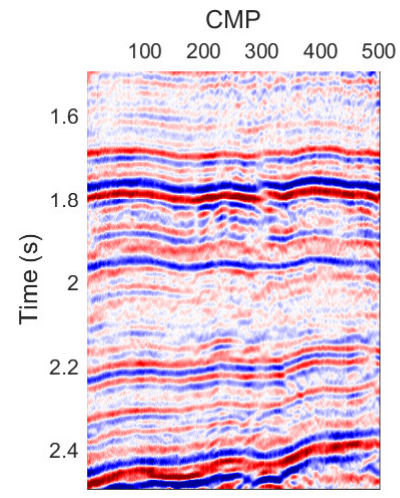}
                \end{minipage}
                \label{fig:real_zoom_A}
        }
        \subfloat[]
        {    
                \begin{minipage}{0.45\linewidth}
                        \centering 
                        \includegraphics[width=0.95\textwidth,height=1\textwidth]{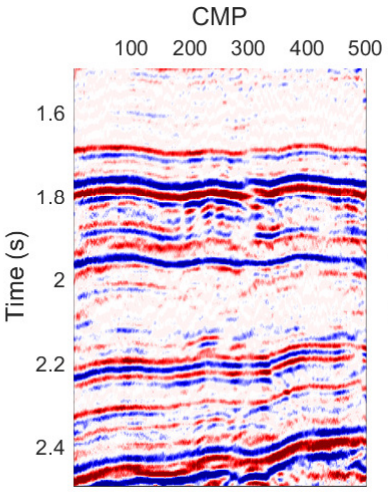}
                \end{minipage}
                \label{fig:real_zoom_B}
        }\\
      \subfloat[]
      {
              \begin{minipage}{0.45\linewidth}
                      \centering 
                      \includegraphics[width=0.95\textwidth,height=1\textwidth]{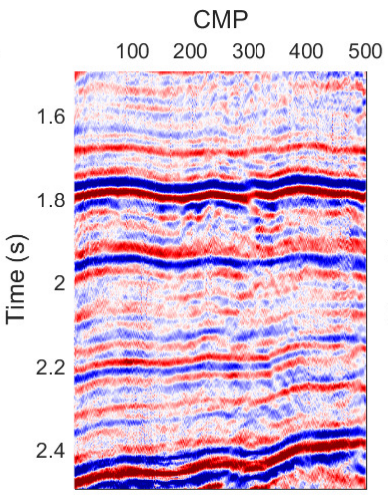}
              \end{minipage} 
              \label{fig:real_zoom_C}
      }
        \subfloat[]
        {
                \begin{minipage}{0.45\linewidth}
                        \centering 
                        \includegraphics[width=0.95\textwidth,height=1\textwidth]{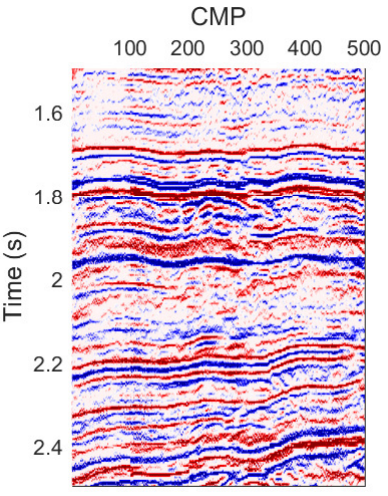}
                \end{minipage}
                \label{fig:real_zoom_D}
        }
        \caption{ (a) Subset of the stacked CMP data  for $1.5\leq t\leq 2.5$\,s and 500 traces.  Deconvolution results for (b) SMBD, (c) F-SMBD, (d) SMBD-SPG (5 iterations).}
        \label{fig:real2}
\end{figure*}
\begin{figure}[htbp]
        \centering
        \subfloat[]
        {
                \begin{minipage}{0.5\linewidth}
                        \centering
                        \includegraphics[width=\textwidth]{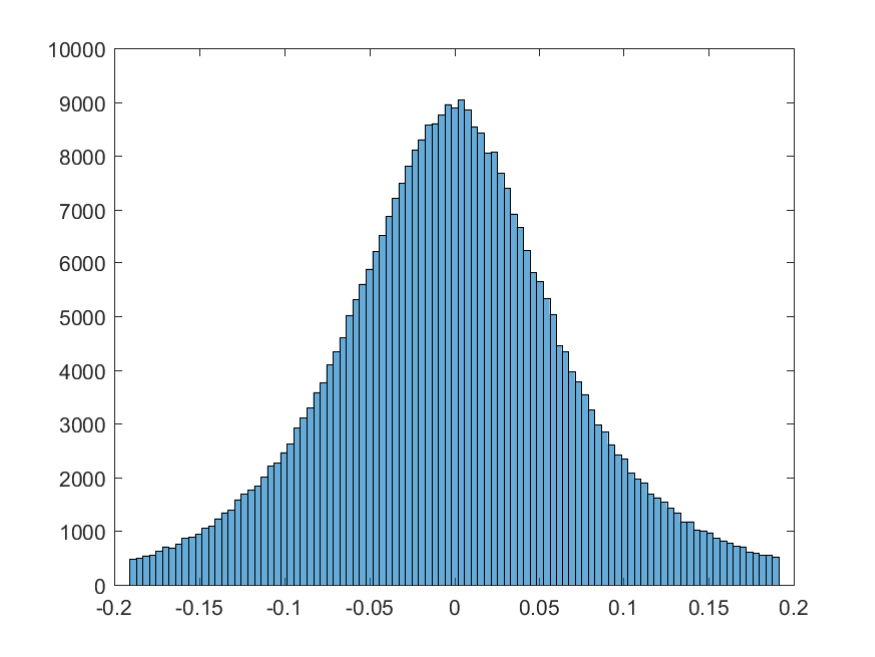}
                \end{minipage}
                \label{fig:histoData}
        }
        \subfloat[]
        {    
                \begin{minipage}{0.5\linewidth}
                        \centering 
                        \includegraphics[width=\textwidth]{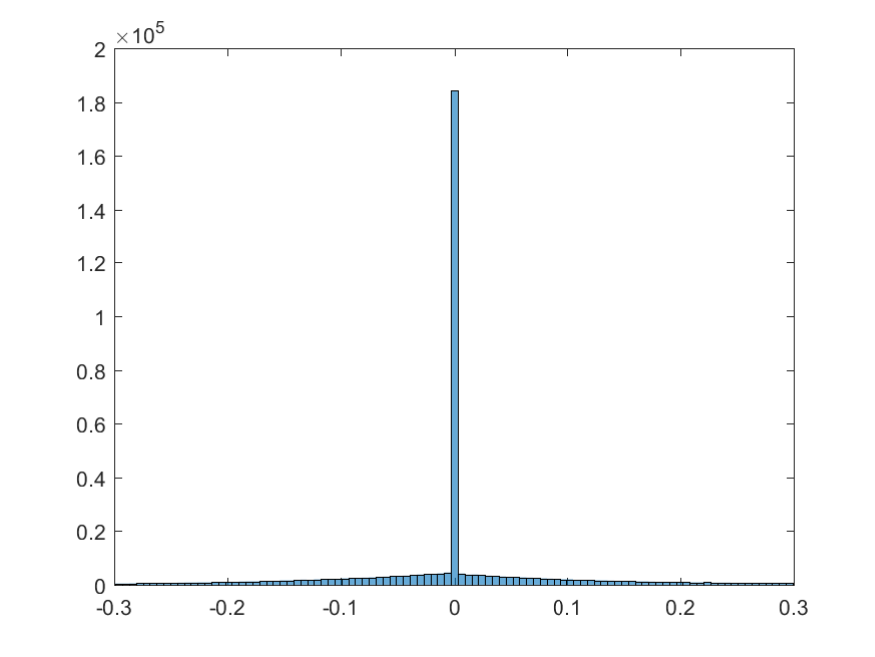}
                \end{minipage}
                \label{fig:histoK}
        }\\
      \subfloat[]
      {
              \begin{minipage}{0.5\linewidth}
                      \centering 
                      \includegraphics[width=\textwidth]{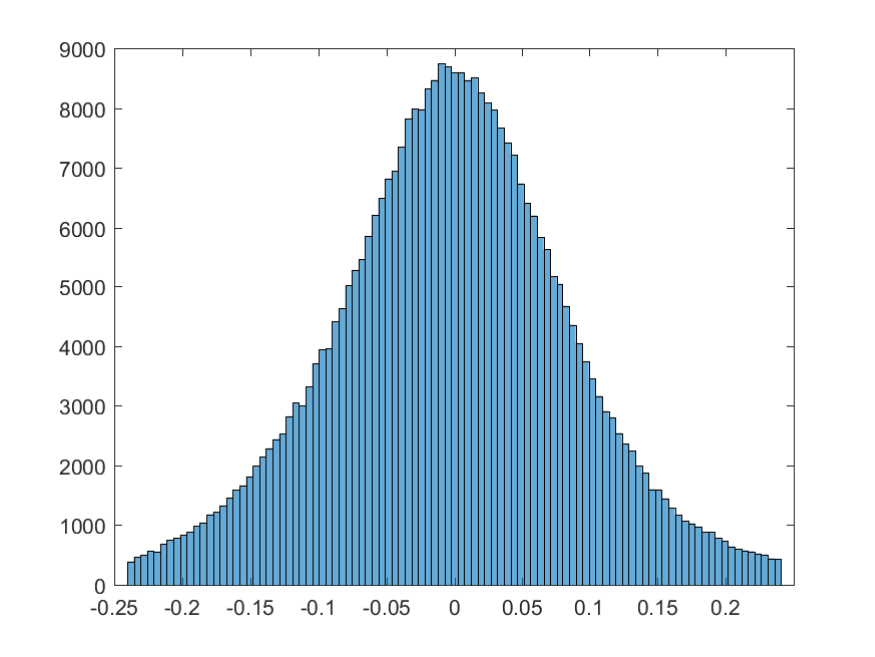}
              \end{minipage} 
              \label{fig:histoF}
      }
        \subfloat[]
        {
                \begin{minipage}{0.5\linewidth}
                        \centering 
                        \includegraphics[width=\textwidth]{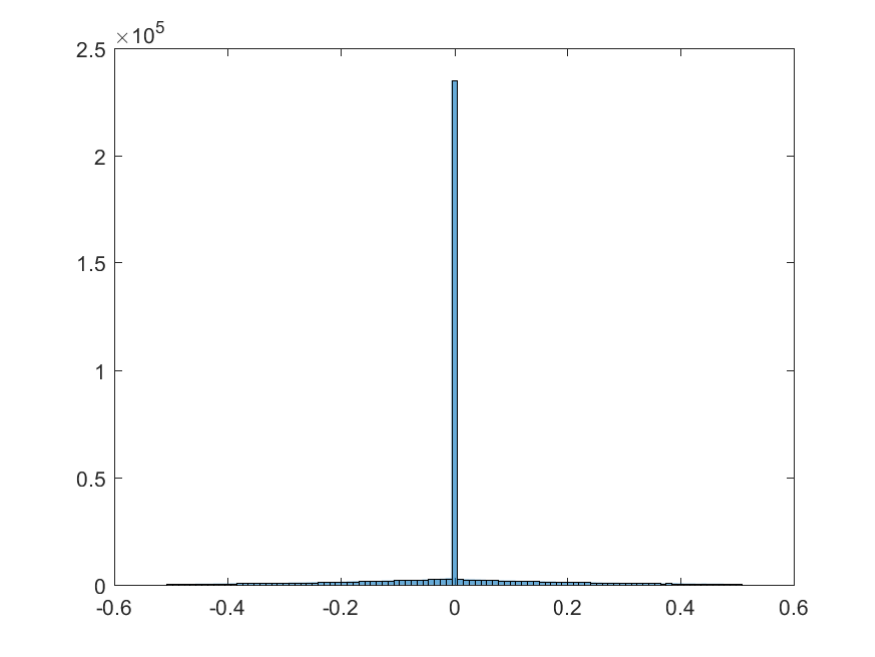}
                \end{minipage}
                \label{fig:histoN}
        }
        \caption{ Histogram of normalized (a) stacked CMP data, and for deconvolution results of (b) SMBD, (c) F-SMBD, (d) SMBD-SPG (5 iterations).}
        \label{fig:histogram}
\end{figure}
\begin{figure*}[htbp]
        \begin{center}
                \includegraphics[width=0.9\textwidth]{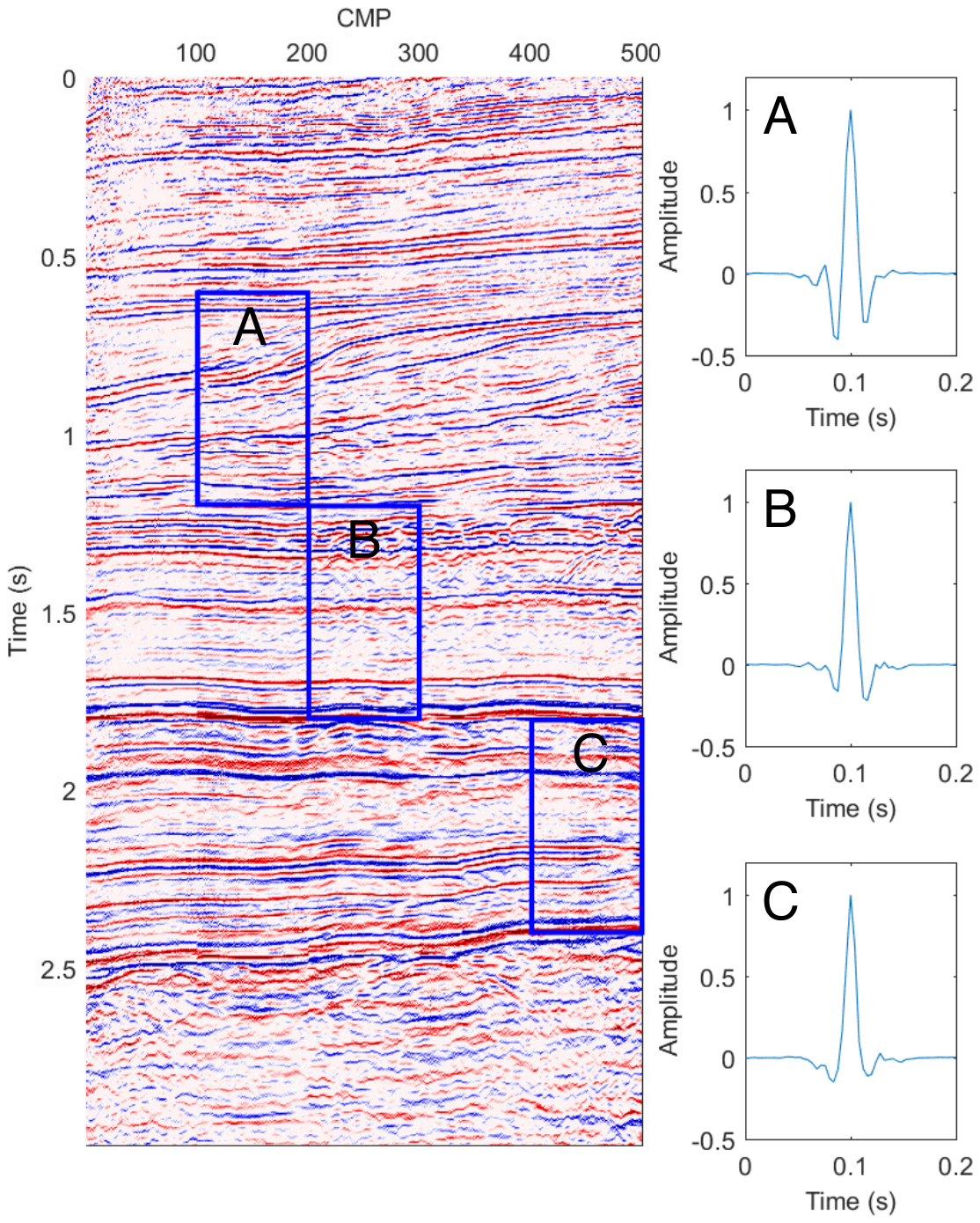}
                \caption{ Recovered wavelet in three sample processing blocks for SMBD-SPG.}
                \label{fig:wave_patch}
        \end{center}
\end{figure*}

Figure \ref{fig:real1}  shows the input data and the  deconvolution results for SMBD-SPG. For comparison, the details of a zoomed-in seismic section before and after deconvolution are shown in Figure \ref{fig:real2} for all methods. It is clear from these results that the proposed algorithm has a more spiky deconvolution output and more weak reflections are preserved than the other two algorithms. In Figure \ref{fig:wave_patch} we show three processing blocks 
in blue rectangles and their corresponding recovered source wavelets, where the consistency and quality of the estimates is easy to observe.
Since the deconvolution results are approximation of the reflectivity, they actually have majority of values around zero. If we show the normalized deconvolution results in the range $[-1,1]$, most of the values are covered by a tiny portion of the color range. This generates a image of large amount of very light color pixels, which makes the visibility bad and ruins many meaningful details. Therefore, we use a zero centered interval $[-m,m]$ to include 95\% of normalized values in the histogram and show the cropped histogram in Figure \ref{fig:histogram}.
        All values goes beyond the interval $[-m,m]$ are pulled back to the edge points of it, and the values inside the interval are intact. Thus, the colorbar we used in Figure \ref{fig:real1} and \ref{fig:real2} applied on a smaller range of values, which enriches visual effects with more details.

\begin{table}[bh!]
        \centering
\caption{Comparison of the computational complexity of SMBD, F-SMBD, and SMBD-SPG.  }
        \label{table_conv}
\begin{tabular}{ll}
\hline \hline
Method & Complexity  \\ \hline
SMBD & $\mathcal{O}$($n^3$)  \\
F-SMBD & $\mathcal{O}$($n^2$) \\
SMBD-SPG & $\mathcal{O}$($n$ log $n$) \\
\hline \hline
\end{tabular}
\end{table}

\section{Conclusions}
In the multichannel blind deconvolution problem, the assumption of an identical seismic source wavelet in all channels leads naturally to a two-step algorithm that alternately estimates the wavelet given the reflectivity and then updates the reflectivity given the wavelet.
Previous multichannel blind deconvolution methods have used the identical wavelet assumption to obtain the Euclid deconvolution property in equation \eqref{eq:matrixXR} which eliminates the wavelet. However, we use all the channels at once to recover the wavelet in the frequency domain which effectively increases the SNR of the recovered wavelet.
For the reflectivity update we exploit sparsity in order to express the reflectivity update as basis pursuit denoising.
The reflectivity estimate is updated for \emph{all channels} via sparse recovery with BPDN, which can be efficiently solved using the SGPL1 package---one of the best available fast $\ell_1$ methods. This approach ensures the computational efficiency of SMBD-SPG.
Furthermore, in SMBD-SPG only two parameters must be set, which makes the scheme easy to implement for real applications. 
As a final comment, we note that the Euclid deconvolution property leads to a term $\|\bA\bx\|_2$ in the SMBD objective functionals in equation \eqref{obj1} and \eqref{eq:modSMBDargmin}, which cannot be incorporated into the BPDN framework and its efficient algorithmic solution.

In the simulation with synthetic data, the quality of the recovery is evaluated versus the known true reflectivity. The SMBD-SPG method is robust to noise, i.e., it provides results better than SMBD and F-SMBD with respect to the normalized correlation and quality metric for a wide range of SNR. 
In addition, to achieve these better quality deconvolution outputs, the computation time for SMBD-SPG is significantly less than SMBD and F-SMBD (more than two orders of magnitude faster in some cases in Figure \ref{fig:mean and std of nets}d).
\section{Appendix: SPG}
\begin{figure}
        \centering
        \includegraphics[width=0.45\textwidth]{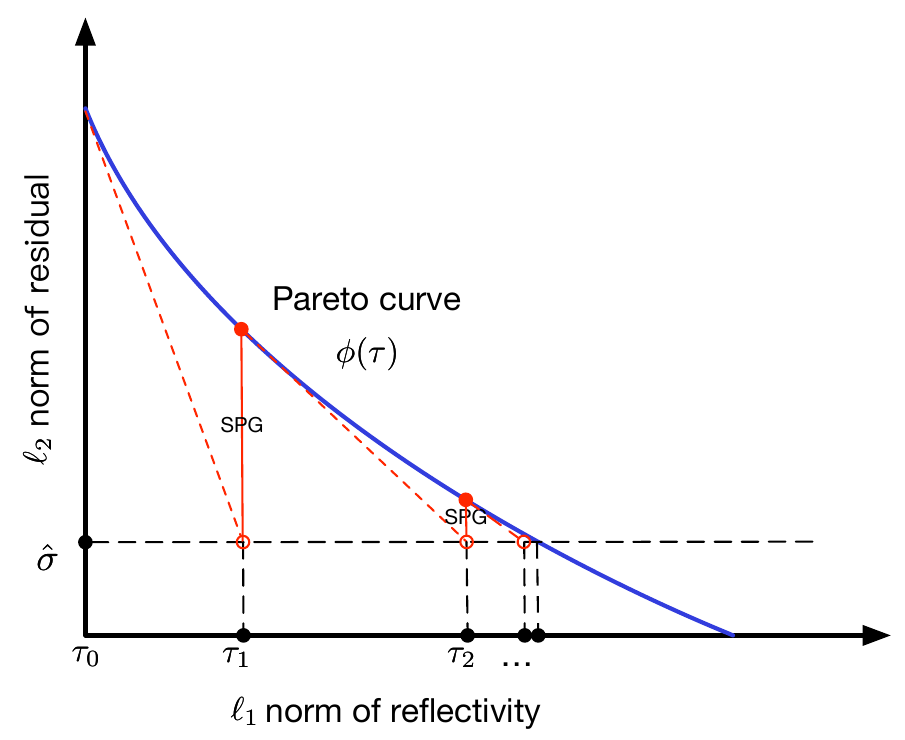}
        \caption{Pareto curve and first three iterations of SPG}
        \label{fig:Pcurve}
\end{figure}

The SPG method employed in this paper, which provides a very efficient numerical solution to equation \eqref{cbyc}, is based on the general idea of sparsity promoting least squares optimization, for details see \cite{vanFri2008,HenvdBAO2008,spgl1:2007}.
In basis pursuit the $\ell_1$ norm is minimized subject to an $\ell_2$ constraint
\begin{equation}\label{bpdn}
\min_\bx \|\bx\|_1 \quad \mbox{subject to}\quad \|\bH\bx-\bb\|_2\le \sigma
\end{equation}
where $\sigma$ is the RMS error in matching the noisy measurements $\bb$ with the model $\bH\bx$.
For any $\sigma\geq 0$ the basis pursuit problem (also known as BPDN if $\sigma>0$) has an equivalent LASSO problem \cite{Tib1996,vanFri2008}
\begin{eqnarray}\label{lasso}
\min_\bx \|\bH\bx-\bb\|_2 \quad \mbox{subject to}\quad \|\bx\|_1\le \tau.
\end{eqnarray}
In other words, after solving \eqref{lasso} for a given $\tau$, the minimum value of $ \|\bH\bx-\bb\|_2$ provides the value of $\sigma$ that would be needed in \eqref{bpdn} to get the same $\bx$ with the smallest $\|\bx\|_1$.
The set of all $(\sigma,\,\tau)$ pairs in this equivalence implicitly define a function $\phi(\tau) =\sigma$, which is convex and differentiable.  
The graph of a typical $\phi(\tau)$ is the Pareto curve shown in Figure \ref{fig:Pcurve}.

We want to solve the BPDN problem in \eqref{bpdn}, but it is more efficient to solve a sequence of LASSO problems \eqref{lasso} to get the BPDN solution.
One catch is that we know $\sigma$, but we don't know $\tau$.
Thus we must generate a sequence of $\tau_k \rightarrow \hat{\tau}$, where $\tau_{k+1} =\tau_k +\Delta\tau_k$, for which \eqref{lasso} yields a sequence of solutions $\{\bx_k\}$ with $\ell_2$ error $\hat\sigma_k$.
For this $\hat{\sigma}_k$, the update of $\tau_k$ is based on a straight line extrapolation using the derivative of the Pareto curve at $(\tau_k,\sigma_k)$: thus, $\Delta\tau_k=(\hat\sigma_k-\phi(\tau_k))/\phi'(\tau_k)$ where $\phi'(\tau_k)=-\|\bH^H(\bH\bx_k-\bb)\|_\infty/\phi(\tau_k)$. 
During the SPG method we (approximately) evaluate $\sigma_k = \phi(\tau_k) =\|\bH\bx_k-\bb\|_2$ whenever we solve \eqref{lasso}, which yields the sequence of filled red dots on the Pareto curve in Figure \ref{fig:Pcurve}. 
The convergence rate of this approach is superlinear which is much faster than that of conventional steepest descent gradient methods which is used in existing schemes such as SMBD and F-SMBD.
{\color{black} As an final comment, we note that including the Euclid deconvolution term $\|\bA \bx\|_2^2$ in the objective functional, as in \eqref{obj1} for SMBD or in \eqref{eq:modSMBDargmin} for modified SMBD,
would destroy the computational simplicity of the SPG algorithm that solves BPDN.}

\section{Acknowledgement}
This work is supported by the Center for Energy and Geo Processing (CeGP) at Georgia Tech. We also appreciate the support provided by the CeGP at King Fahd University of Petroleum and Minerals (KFUPM) under project number GTEC1311.

\bibliographystyle{IEEEtran}
\bibliography{refs}

\end{document}